\documentclass[12pt,a4paper]{iopart}
\usepackage{graphicx}
\usepackage[utf8]{inputenc} 
\usepackage{color}
\usepackage{soul} 
\usepackage[
,textwidth=15cm
,textheight=20cm
,verbose
,dvips
]{geometry}
\usepackage[numbers,comma,square,sort&compress,merge]{natbib}

\begin{document}

\title[Properties of ZnO nanowires grown on Si(001) by chemical vapour transport]{Coalescence, crystallographic orientation and luminescence of ZnO nanowires grown on Si(001) by chemical vapour transport}

\author{S Fernández-Garrido$^{1,*}$, C Pisador$^{1}$, J~Lähnemann$^{2}$,\\ S Lazić$^{3}$, A~Ruiz$^{4}$, and A Redondo-Cubero$^{1}$}
\address{$^{1}$Grupo de Electrónica y Semiconductores, Dpto. Física Aplicada,
Universidad Autónoma de Madrid, C/ Francisco Tomás y Valiente 7, 28049
Madrid, Spain}
\address{$^{2}$Paul-Drude-Institut für Festkörperelektronik, Leibniz-Institut im Forschungsverbund Berlin e.V., 
Hausvogteiplatz 5–7, 10117 Berlin, Germany}
\address{$^{3}$Universidad Autónoma de Madrid (UAM), Instituto Universitario de Ciencia de Materiales “Nicolás Cabrera” (INC) and Condensed Matter Physics Center (IFIMAC), Departamento de Física de Materiales, 28049, Madrid, Spain}
\address{$^{4}$Instituto de Ciencia de Materiales de Madrid, CSIC, C/ Sor Juana Inés de la Cruz 3, Cantoblanco, 28049, Madrid, Spain}
\ead{$^{*}$sergio.fernandezg@uam.es}

\begin{abstract}
We analyse the morphological, structural and luminescence properties of self-assembled ZnO nanowires grown by chemical vapour transport on Si(001). The examination of nanowire ensembles by scanning electron microscopy reveals that a non-negligible fraction of nanowires merge together forming coalesced aggregates during growth. We show that the coalescence degree can be unambiguously quantified by a statistical analysis of the cross-sectional shape of the nanowires. The examination of the structural properties by X-ray diffraction evidences that the nanowires crystallize in the wurtzite phase, elongate along the \textit{c}-axis, and are randomly oriented in plane. The luminescence of the ZnO nanowires, investigated by photoluminescence and cathodoluminescence spectroscopies, is characterized by two bands, the near-band-edge emission and the characteristic defect-related green luminescence of ZnO. The cross-correlation of scanning electron micrographs and monochromatic cathodoluminescence intensity maps reveals that: (i) coalescence joints act as a source of non-radiative recombination, and (ii) the luminescence of ZnO nanowires is inhomogeneously distributed at the single nanowire level. Specifically, the near-band-edge emission arises from the nanowire cores, while the defect-related green luminescence originates from the volume close to the nanowire sidewalls. Two-dimensional simulations of the optical guided modes supported by ZnO nanowires allow us to exclude waveguiding effects as the underlying reason for the luminescence inhomogeneities. We thus attribute this observation to the formation of a core-shell structure in which the shell is characterized by a high concentration of green-emitting radiative point defects with respect to the core.
\end{abstract}



\maketitle

\section{Introduction}
The synthesis of semiconductors in the form of nanowires (NWs) instead of epitaxial films provides new degrees of freedom for the design of opto- and electronic devices as well as the possibility of combining otherwise incompatible materials \cite{Patolsky2005,Li2006,Yan2009}. Among the wide variety of semiconductors that can be grown as NWs, ZnO and GaN stand out. Regardless of the growth technique, these two wide band-gap compound semiconductors, which typically crystallize in the wurtzite phase, exhibit a pronounced tendency to spontaneously form single-crystalline NWs on a diverse variety of substrates \cite{Yi2005,Schmidt-mende_2007,Xu2011,Consonni_PSSRL_2013}, with astonishing examples such as metal foils and concrete \cite{Calabrese_APL_2016,LePivert2019,Campos2020}. More specifically, in contrast to other semiconductor materials, the formation of ZnO and GaN NWs does not require the use of metal particles (in the literature referred to as catalysts) to induce their uniaxial growth by means of the vapour-liquid-solid growth approach \cite{Wagner_apl_1964}. Instead, for these two material systems, NWs spontaneously form under specific growth conditions. In the particular case of ZnO, the synthesis of this compound as NWs is the subject of worldwide research because of their multiple potential applications, including the development of photocatalytic water splitting cells, piezoelectric nanogenerators, gas and biological sensors, ultraviolet detectors, electrochromic displays, field effect transistors, and light emitters \cite{Yi2005,Schmidt-mende_2007,Xu2011,Comini2016,Shanmugam2017}.   

The catalyst-free self-assembled formation of ZnO and GaN NWs is as attractive as popular because it is the simplest method to obtain nanostructures with a structural perfection comparable to that of bulk materials \cite{Djurisic2006,Reshchikov2009,Zettler_CGD_2015}. However, this growth approach also has inherent limitations and drawbacks. Besides the poor degree of control over the NW radius and areal density, a non-negligible fraction of NWs typically merge during growth forming coalesced NW aggregates. As a result of the mutual misorientation of coalesced NWs, the coalescence process can potentially deteriorate the structural quality of the NWs by the creation of dislocated tilt and twist boundaries at the coalescence joints. In the case of GaN, the physical origin of coalescence and the impact of this undesired effect on the NW structural and optical properties have been investigated in depth \cite{Consonni_APL_2011_2,Jenichen_NT_2011,Fan2014,Brandt_CGD_2014,Fernandez-Garrido_NT_2014,Kaganer_NL_2016,Auzelle2016,Treeck2018}. Specifically, it was demonstrated that the coalescence process leads indeed to the formation of dislocation networks \cite{Consonni2009,Jenichen_NT_2011,Grossklaus_JCG_2013,Fan2014}, which act as a source of non-radiative recombination, as well as to the introduction of both inhomogeneous \cite{Jenichen_NT_2011,Fernandez-Garrido_NT_2014} and homogeneous strain \cite{Auzelle2016}. NW coalescence was also recognized in the case of ZnO, but has received much less attention \cite{Jeong2005,Han2006,Liu2008,Kumar2013,Zhu2015}. Particularly, no attempts were made to unambiguously quantify the coalescence degree or to systematically correlate this effect with the structural perfection and luminescence properties of NW ensembles.

In this work, we present a comprehensive characterization of self-assembled ZnO NWs grown by chemical vapour transport (CVT) on Si(001), as required for the monolithic integration of ZnO devices with the complementary metal oxide semiconductor (CMOS) technology. We devote a special attention to the analysis of the coalescence degree, which is unambiguously determined from a mathematical point-of-view by examining the cross-sectional shape of the NWs according to the statistical methods proposed by Brandt et al.~\cite{Brandt_CGD_2014} for the analysis of the shape of one-dimensional nanostructures. The study of the luminescence properties by photoluminescence (PL) and cathodoluminescence (CL) spectroscopies evidences that coalescence joints introduce non-radiative recombination as well as an inhomogeneous luminescence distribution at the single NW level. The latter effect, attributed to a higher concentration of green-emitting radiative point defects at the NW sidewalls, results in the formation of a core-shell structure in which the core and the shell emit light a different wavelengths.

\section{Experiment}

The ZnO sample under investigation is grown by CVT on a $2\times$1~cm$^{2}$ Si(001) substrate. To enhance the nucleation of ZnO NWs, a Zn layer is deposited prior to ZnO growth on the as-received Si surface \cite{Fu1998}. This layer of Zn is deposited by radio-frequency magnetron sputtering in an Alcatel A450 system equipped with a pure (5N) Zn target. The sputtering process takes place at a pressure of $10^{-2}$~mbar, and is performed using a flow of 50~sccm of pure (5N) Ar and a rf-power of 50~W. The sputtering time is 15~s, which results in a layer thickness of 6~nm. The growth of ZnO by CVT is carried out in a horizontal quartz tube furnace using pure (5N5) Zn powder and (5N) O$_{2}$ as precursors, and (5N) Ar as carrier gas. More details about the experimental system can be found elsewhere \cite{GarciaNunez2014}. To avoid the oxidation of the Zn layer during the heating of the furnace, the tube is heated under a constant Ar flow of 140~sccm. Upon reaching the growth temperature, namely, 900~$^{\circ}$C, we introduce an O$_{2}$ flow of 10~sccm to initiate the growth. The growth time is 4~hours. After the growth, the sample is cooled down by natural convection. 

The morphological and structural properties of the sample are investigated by scanning electron microscopy (SEM) and high resolution X-ray diffraction (HRXRD), respectively. Scanning electron micrographs are acquired employing a Zeiss Ultra55 field-emission microscope. X-ray diffraction experiments are performed with Cu$K\alpha_{1}$ radiation (wavelength $\lambda=1.54056$~\AA) using a 4-axis Bruker AXS D8 Advance diffractometer equipped with a Göbel mirror and a four-bounce Ge$(220)$ monochromator. Depending on the type of measurement, we use either a Nal scintillation point detector or a one-dimensional position-sensitive Si strip detector (Lynxeye). To acquire the symmetric $\omega/2\theta$ scan, a 1~mm slit is placed in front of the point detector. The $\omega$ scan across the ZnO~$0002$ Bragg reflection and the  ZnO~$\lbrace10\bar{1}1\rbrace$ pole figure are both recorded with the point detector using a 3~mm slit. The reciprocal space maps across the Si~004 and ZnO~0004 Bragg reflections are collected with fixed counting times using the one-dimensional position sensitive detector at sequential $\omega$ angles. 

The luminescence properties of the sample are investigated by low-temperature (11~K) $\mu$-PL and room-temperature CL. CL measurements are carried out at an acceleration voltage of 5~kV using a Gatan Mono-CL4 system equipped with a parabolic mirror for light collection and with both a photomultiplier and a charge-coupled device (CCD) for detection. The system is mounted to the Zeiss Ultra55 microscope used for imaging. Monochromatic images are taken with a wide spectral resolution window of 25~nm. Spectra are recorded with a reduced spectral window of 6~nm. The time-integrated $\mu$-PL experiments are performed on a sample mounted in a He flow cryostat. A continuous-wave (cw) He-Cd laser (Kimmon) operating at 325~nm is used as excitation source. The laser beam is focused onto a spot of about $1.5$~$\mu$m diameter on the sample surface through a $50\times$ microscope objective lens (Mitutoyo M Plan UV) with a numerical aperture of 0.41. The excitation power density is 805~W/cm$^{2}$. The emitted light is collected by the same objective lens, dispersed by a single-grating monochromator (Horiba/Jobin Yvon 1000M Series II, spectral resolution of $\approx350$~$\mu$eV) and detected with a liquid-N-cooled CCD camera (Symphony II from Horiba Scientific). To prevent the excitation laser from entering the monochromator, a 325 nm RazorEdge ultrasteep long-pass edge filter (Semrock LP03-325RE-25) is inserted in the optical collection path. 


\section{Results}
\subsection{Morphology and coalescence degree}

\begin{figure*}
\hspace{1in}\includegraphics*[width=0.85\textwidth]{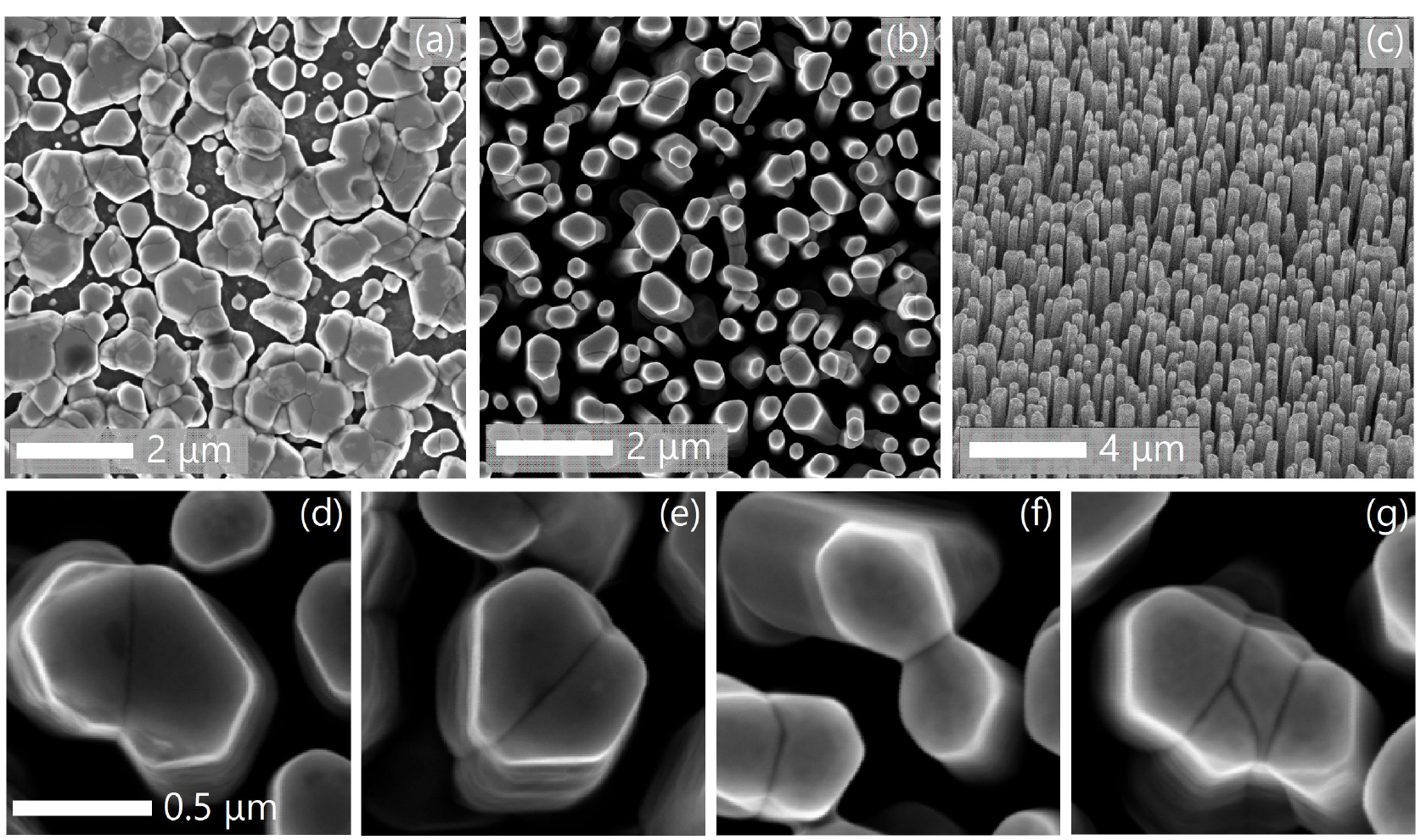}
\caption{\label{SEM} Plan-view scanning electron micrographs of the ZnO (a) pseudo-compact layer and (b) NW ensembles obtained at opposite sides of the sample. (c) Bird's eye-view scanning electron micrograph of the NW ensemble. (d)--(g) Magnified plan-view scanning electron micrographs illustrating the formation of different types of coalesced NW aggregates. The magnification is the same for all micrographs. The micrographs shown in (d) and (f) exemplify NW aggregates with coalesced boundaries parallel to \textit{A}- and \textit{M}-planes, respectively.}
\end{figure*}

We examine the morphological properties of the sample by SEM (Fig.~\ref{SEM}). The study reveals a continuous variation in the morphology across the sample surface that evidences a gradient in the growth parameters. Specifically, from one sample side to the other, we observe a progressive transition from a pseudo-compact layer [Fig.~\ref{SEM}(a)] to a NW ensemble with an areal density of $\approx5\times10^{8}$~cm$^{-2}$ [Figs.~\ref{SEM}(b) and \ref{SEM}(c)]. The progressive transition between these two different regions evidences that the pseudo-compact layer is formed by the coalescence of closely spaced NWs. Additional cross-sectional scanning electron micrographs performed on both sample sides (not shown here) show that the layer is approximately 800~nm thick and the NWs are about 7~$\mu$m long. Hereafter, we constrain our study to these two different sample regions. 

According to the literature \cite{Yi2005,Park2008,Kumar2013}, spontaneously formed ZnO NWs crystallize in the thermodynamically stable wurtzite phase, elongate along the \textit{c}-axis, and exhibit abrupt sidewall facets formed by $\lbrace1\bar{1}00\rbrace$  planes (known as \textit{M}-planes). The cross-sectional shape of our ZnO NWs is compatible with these expectations. As can be observed in Fig.~\ref{SEM}(b), most NWs have a cross-sectional shape close to a regular hexagon, which would be formed by the intersection of six $\lbrace1\bar{1}00\rbrace$ planes. Nevertheless, as illustrated in Figs.~\ref{SEM}(d)--\ref{SEM}(g), there is also a non-negligible number of NWs with a cross-sectional shape that deviates significantly from a regular hexagon. Although variations in the growth rate from facet to facet may distort the ideal hexagonal cross-sectional shape, the clear observation of grain boundaries in Figs.~\ref{SEM}(d)--\ref{SEM}(g) evidences that highly distorted cross-sectional shapes are caused by the formation of coalesced NW aggregates. Interestingly, coalescence joints are sharp, straight and often parallel to either $\lbrace1\bar{1}00\rbrace$ or $\lbrace1\bar{12}0\rbrace$ (\textit{A}-plane) planes, provided that the sidewall facets of the NWs are formed by \textit{M}-planes as stated above.

Due to the potentially detrimental impact of coalescence on the structural perfection of ZnO NWs, it is important to find a way to objectively assess the coalescence degree of dense ensembles, i.\,e., the ratio between coalesced aggregates and the total number of objects. In the following, we analyse the coalescence degree of our ZnO NW ensemble using the two different mathematical methods proposed by Brandt \textit{et al}.~\cite{Brandt_CGD_2014} for one-dimensional nanostructures. Both methods, which are based on the analysis of the cross-sectional shape of the NWs, were previously used to assess the coalescence degree of GaN NW ensembles \cite{Brandt_CGD_2014,Fernandez-Garrido_NT_2014,Kaganer_NL_2016}, but have never been applied to ZnO nanostructures. 

\begin{figure*}
\includegraphics*[width=\textwidth]{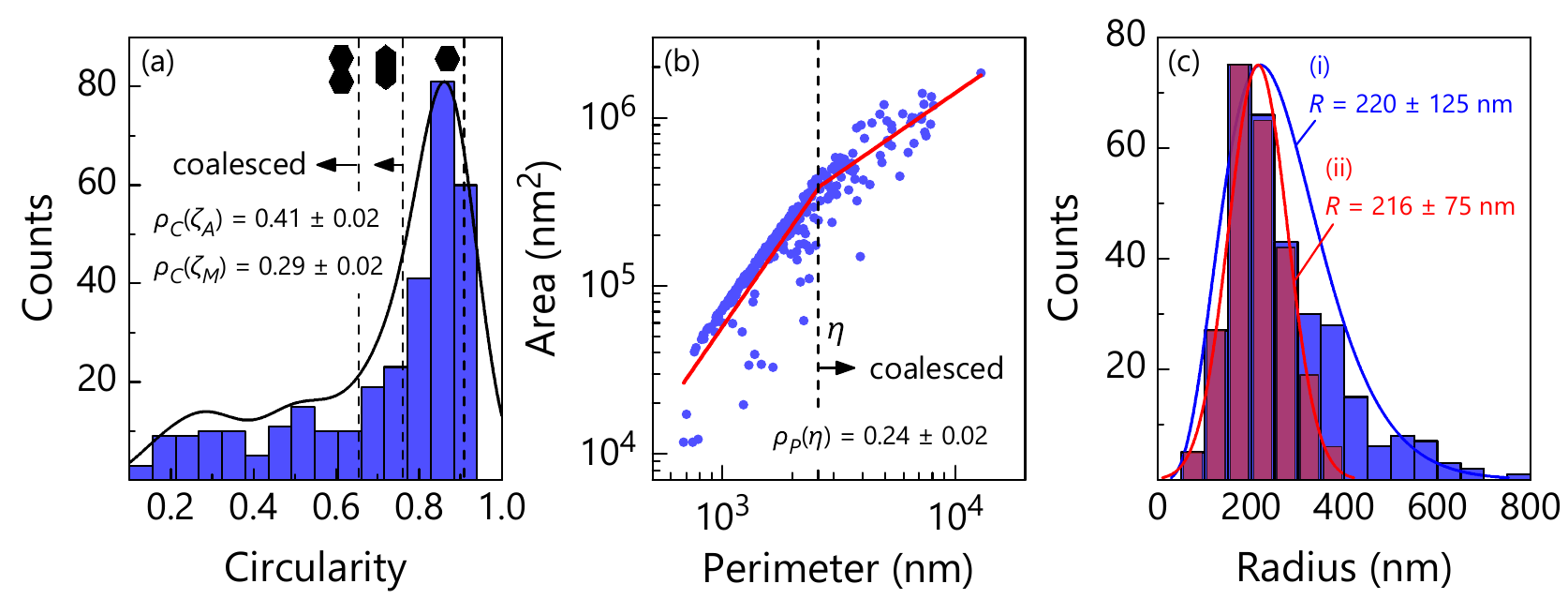}
\caption{\label{histograms}(Colour online) (a) Circularity histogram corresponding to the ZnO NW ensemble. The solid line is a kernel density estimation. The vertical dashed lines indicate the circularity of the geometrical shapes displayed next to them. The coalescence degrees obtained using the criterion of a minimum circularity of either $\zeta_{A}=0.762$ or $\zeta_{M}=0.653$ are also provided in the figure. (b) Area-perimeter plot of the ZnO NW ensemble. The solid line shows a fit of Eq.~(\ref{area vs perimeter}) to the experimental data (solid symbols). The vertical dashed line indicates the critical perimeter $\eta$. The coalescence degree derived from the critical perimeter $\eta$ is shown in the figure. (c) Radius distribution of either (i) all NWs or (ii) only those which are not coalesced according to the area-perimeter plot shown in (b). The histograms are fit by (i) a shifted gamma and (ii) a normal distribution, as shown by the solid lines, yielding the respective mean diameter and FWHM indicated in the figure.}
\end{figure*}

The first method reported in Ref.~\cite{Brandt_CGD_2014} to evaluate the coalescence degree of a NW ensemble consists in the analysis of the circularity of the cross-sectional shape of the NWs. The circularity is defined as 
\begin{equation}
\label{circularity}
C=4\pi{A}/P^{2},
\end{equation}
where $A$ and $P$ are the shape area and perimeter, respectively. Hence, $C=1$ for a circle, $C\approx0.907$ for a regular hexagon, and $C\ll1$ for a strongly elongated object. Since coalescence tends to form objects with rather elongated cross-sectional shapes, a threshold circularity value $\zeta$ can be used to systematically distinguish between single NWs and coalesced aggregates. The coalescence degree is thus given by
\begin{equation}
\label{circularity}
\rho_{C}(\zeta)=N_{C<\zeta}/N,
\end{equation}
where $N_{C<\zeta}$ is the number of NWs with \textit{C} values lower than $\zeta$, and \textit{N} the total number of objects. As explained in detail in Ref.~\cite{Brandt_CGD_2014}, for NWs with an intrinsic hexagonal cross-sectional shape, reasonable values of $\zeta$ are 0.762 and 0.653, depending on whether coalescence boundaries are parallel to either \textit{A}- or \textit{M}-planes, respectively. Figure~\ref{histograms}(a) shows the histogram of the circularity for the NW ensemble, as derived from the analysis of plan-view scanning electron micrographs covering more than 300~NWs using the open-source software ImageJ \cite{Abramoff_Biophotonics_2004}. The histogram peaks at a value close to that of a regular hexagon and has a clear tail toward low circularities. The coalescence degree obtained taking $\zeta_{A}=0.762$ is $0.41\pm0.02$ and the one obtained taking $\zeta_{M}=0.653$ is $0.29\pm0.02$. Therefore, the coalescence degree strongly depends on the chosen value of $\zeta$.

The second method proposed in Ref.~\cite{Brandt_CGD_2014} to quantify the coalescence degree does not require the arbitrary definition of a threshold value, which inevitably introduces an element of ambiguity, as seen above. The method is based on the analysis of the area-perimeter relationship of the cross-sectional shape of the NWs. As described in Ref.~\cite{Brandt_CGD_2014}, the area of equilateral shapes, such as a regular hexagon, is proportional to the square of their perimeter regardless of their circularity. In contrast, for linearly elongated shapes, the area depends linearly on the perimeter. Consequently, the coalescence degree can be extracted from an area-perimeter plot. Figure~\ref{histograms}(b) presents the area-perimeter relationship for the same NWs as those used to construct the histogram shown in Fig.~\ref{histograms}(a). As can be observed, the relationship changes abruptly from quadratic to nearly linear at a certain critical perimeter $\eta$. The coalescence degree can thus be defined as
\begin{equation}
\label{sigma P}
\rho_{P}(\eta)=N_{P>\eta}/N,
\end{equation}
where $N_{P>\eta}$ is the number of NWs with a perimeter larger than $\eta$, and $N$ is again the total number of objects. To obtain the value of $\eta$, we fit the following scaling law to the experimental data:
\begin{equation}
\label{area vs perimeter}
A=\alpha\left[P^{2}H\left(\eta-P\right)+\eta^{2-\beta}P^{\beta}H\left(P-\eta\right)\right],
\end{equation}
where $\alpha$ and $\beta$ are fitting parameters, and \textit{H} is the Heaviside step function \cite{Brandt_CGD_2014}. The fit yields $\alpha=0.0571\pm0.002$, $\beta=0.95\pm0.04$, and $\eta=(2.61\pm0.12)\times10^{3}$~nm. The coalescence degree derived from Eq.(\ref{sigma P}) is $0.24\pm0.02$. This value is far from $\rho(\zeta_{A})=0.41\pm0.02$, but close to $\rho(\zeta_{M})=0.29\pm0.02$. 

To validate our results, we take advantage of the clear observation of coalescence joints to manually count the total number of coalesced objects. The coalescence degree obtained from the manual inspection of the same plan-view micrographs as those used to assess $\rho_{C}(\zeta)$ and $\rho_{P}(\eta)$ is 0.28. This value compares quite well with $\rho_{C}(\zeta_{M})$ and $\rho_{P}(\eta)$. In consequence, the manual analysis of the micrographs validates the statistical methods proposed by Brandt \textit{et al.}~\cite{Brandt_CGD_2014} to quantify the coalescence degree of ZnO NW ensembles, and evidences that $\zeta_{M}$ is the most suitable threshold to derive the coalescence degree using the criterion of a minimum circularity.

To conclude the analysis of the morphological properties of our ZnO NW ensemble, we determine and analyse the NW radius distribution paying special attention to the impact of coalescence. Figure~\ref{histograms}(c) shows the distribution of the equivalent-disk radius $R=(A/\pi)^{1/2}$ including either (i) all NWs or (ii) only those which are not coalesced according to the area-perimeter plot. The former one is clearly skewed, tailing toward larger radii, and well described by a shifted gamma distribution \cite{Brandt_CGD_2014} centred at 220~nm with a full width at half maximum (FWHM) of 125~nm. In contrast, the latter one is symmetric and well described by a normal distribution centred at 216~nm with a FWHM of 75~nm. Therefore, as in the case of GaN NWs \cite{Brandt_CGD_2014}, the asymmetry of the radius distribution of all objects seen in plan-view micrographs is caused by the formation of coalesced aggregates.

\subsection{Crystallographic orientation}\label{X-ray}
\begin{figure*}
\includegraphics*[width=0.9\textwidth]{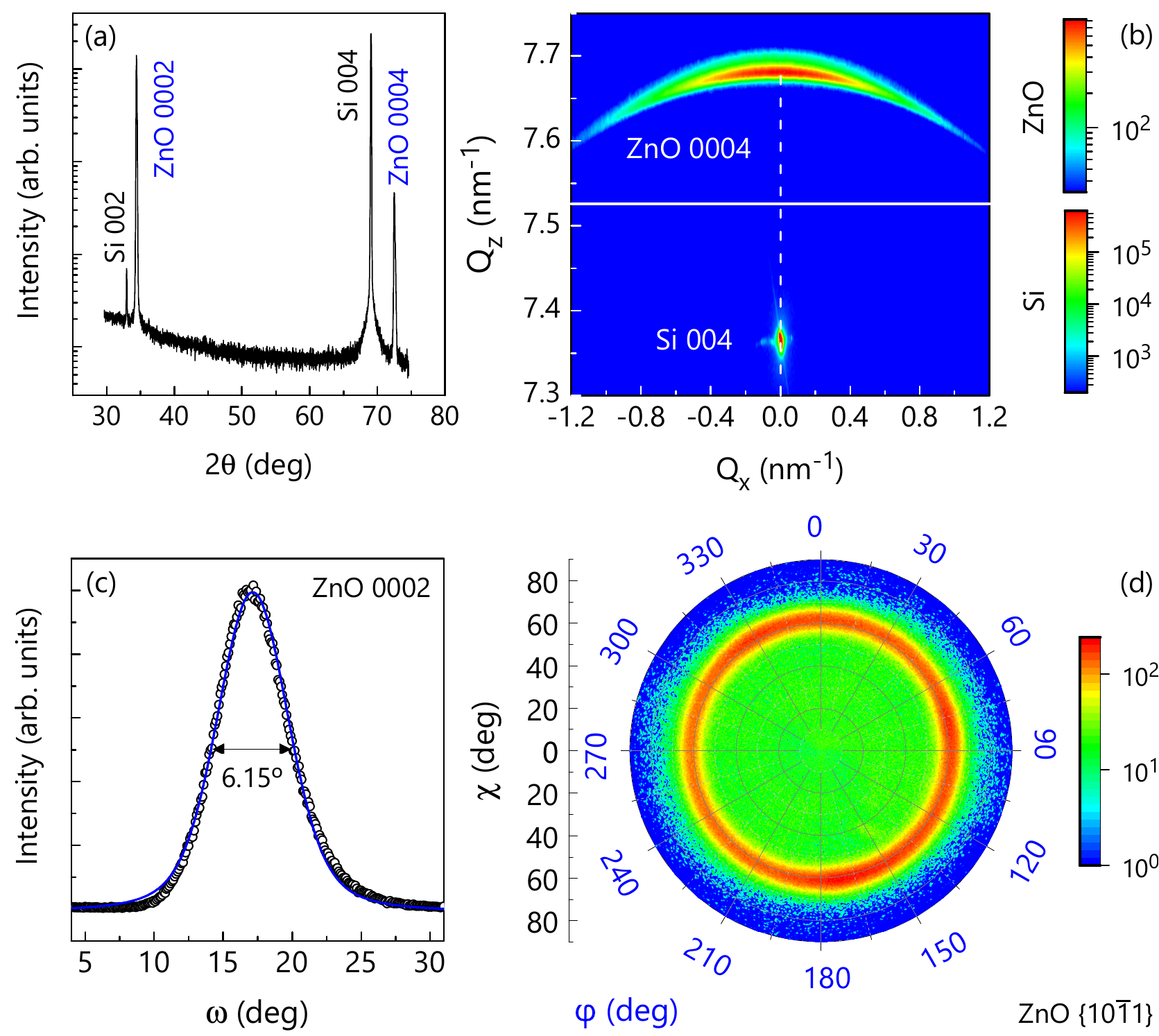}
\caption{\label{fig:XRD} (Colour online) XRD analysis of the ZnO NW ensemble. (a) Symmetric $\theta/2\theta$ scan. (b) Reciprocal space maps around the $004$ and $0004$ Bragg reflections of the Si(001) substrate and the ZnO NWs, respectively. (c) $\omega$ scan across the symmetric ZnO $0002$ Bragg reflection. The experimental data are the open circles and the solid line represents a fit using a Voigt function. (d) ZnO $\lbrace10\bar{1}1\rbrace$ pole figure. In (b) and (d), the X-ray intensity is represented in arbitrary units using a logarithmic colour scale, as indicated by the corresponding scale bars shown next to the plots.}
\end{figure*}

To elucidate the crystallographic properties of the ZnO NWs, we analyse the NW ensemble by XRD. Figure~\ref{fig:XRD}(a) shows a symmetric $\theta$/2$\theta$ scan performed to examine both the crystal phase and the out-of-plane orientation of the ZnO NWs. Besides the peaks related to the Si(001) substrate, we detect the $0002$ and $0004$ Bragg reflections of ZnO. The lack of any additional ZnO-related peak confirms that, as previously assumed, our ZnO NWs crystallize in the wurtzite modification and elongate along the \textit{c}-axis. The reciprocal space map shown in Fig.~\ref{fig:XRD}(b), acquired around the $004$ and $0004$ Bragg reflections of Si and ZnO, respectively, provides a further evidence that ZnO NWs are oriented along the \textit{c}-axis and perpendicular to the substrate surface (note that the Si and ZnO reflections are vertically aligned). The 0004 Bragg reflection of ZnO is, however, not a spot but a wide arc, revealing the presence of misoriented NWs. To assess the out-of-plane orientation distribution of the ZnO NWs, i.\,e., the tilt, we performed the $\omega$ scan across the ZnO 0002 Bragg reflection shown in Fig.~\ref{fig:XRD}(c). The width of the tilt distribution, estimated from the FHWM of the peak, is $6.15^{\circ}$.  

In order to investigate the in-plane epitaxial relationship between the ZnO NWs and the Si substrate, we record the ZnO $\lbrace10\bar{1}1\rbrace$ pole figure depicted in Fig.~\ref{fig:XRD}(d). As can be seen, we do not observe diffraction spots but a ring. The ZnO NWs are thus randomly oriented in plane. This result is attributed to the native amorphous SiO$_{2}$ layer present at the substrate surface prior to the growth. 

\subsection{Luminescence}
 
\begin{figure}
\hspace{1in}\includegraphics[width=0.45\textwidth]{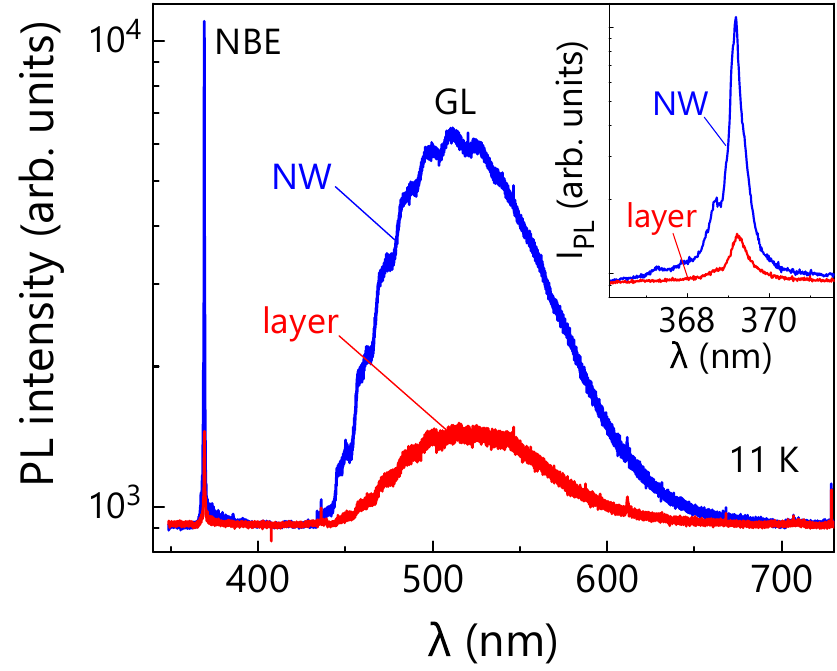}
\caption{\label{fig:LTPL}(Colour online) Low temperature (11~K) $\mu$-PL spectra of the pseudo-compact layer (red line) and the NW ensemble (blue line). The luminescence extends from the near-band-edge (NBE) region to the green spectral range, where we detect the characteristic defect-related green luminescence (GL) of ZnO. The inset shows the NBE luminescence in more detail. As in the main figure, the PL intensity is shown on a logarithmic scale.}
\end{figure}

We investigate the luminescence properties of the sample by PL and CL. Figure~\ref{fig:LTPL} shows the low-temperature (11~K) $\mu$-PL spectra of the pseudo-compact layer and the NW ensemble. Regardless of the sample morphology, we observe a strong near-band-edge UV emission in a narrow wavelength range between 366 and 371 nm and a broad band centred at about 515~nm. The UV luminescence, which is shown in more detail in the inset of Fig.~\ref{fig:LTPL}, is comprised of several PL lines originating from the recombination of excitons bound to different neutral or ionized donors \cite{Meyer2004,Oezguer2006,Djurisic2006}. 
Meanwhile, the broad green emission around 515~nm, consisting of a zero-phonon line at about $2.85$~eV and nine longitudinal-optical phonon replicas separated by $\approx70$~meV, exhibits a characteristic fine structure with clear doublet features split by $\approx27$~meV. According to the literature, a fine-structured green band, as observed here, is caused by the presence of either deep acceptor states related to isolated Zn vacancies V$_\mathrm{Zn}$ \cite{ZYao2015} or Cu$_\mathrm{Zn}$ defects \cite{Dingle1969,Oezguer2006,Reshchikov2007,Katiyar2013}. Since Cu is a common impurity in Zn powder, with concentrations in the order of several parts per million, the presence of this element in our sample is certainly plausible.

The comparison of the PL spectra of the pseudo-compact layer and the NW ensemble brings to light a clear difference. The luminescence from the NW ensemble is much more intense than the one from the layer. In particular, the near-band-edge luminescence is one order of magnitude brighter. We explain this result in terms of both an improved light extraction/collection efficiency \cite{Park2010,Djavid2016,Hauswald2017,Lin2018} and a lower concentration of non-radiative recombination centres in the NW case, as will be shown below by spatially resolved CL measurements. 

\begin{figure}
\hspace{1in}\includegraphics[width=0.45\textwidth] {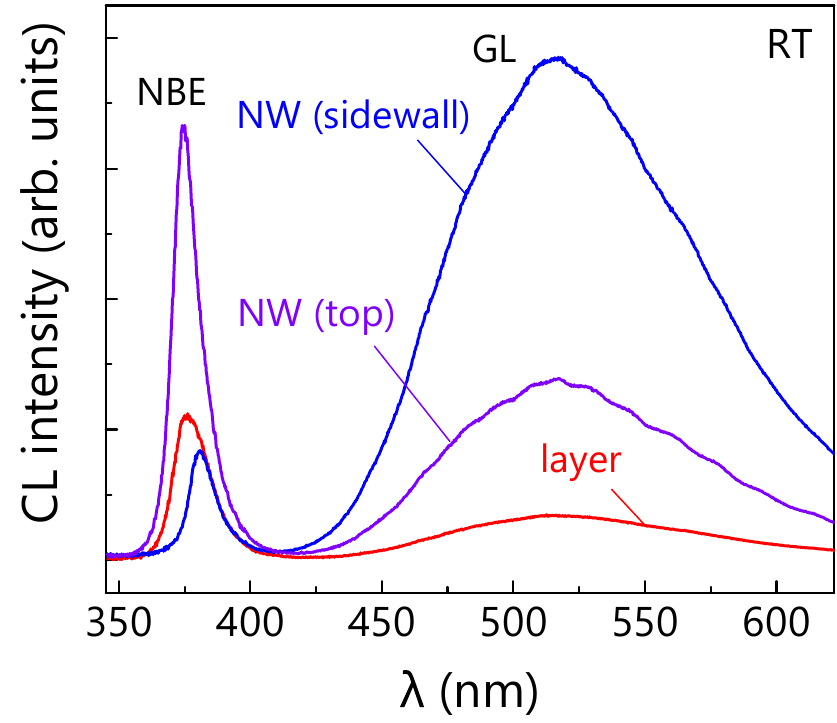} 
\caption{\label{fig:CL}(Colour online) Room-temperature CL spectra of the pseudo-compact layer (red line) and the NW ensemble (violet and blue lines). The spectra of the NW ensemble are acquired at two different locations, namely, on a NW sidewall (blue line) and on a NW top facet (violet line).}
\end{figure}

Figure~\ref{fig:CL} presents the room-temperature CL spectra of the pseudo-compact layer and the NW ensemble. In the case of the NW ensemble, we present two spectra acquired at different locations, namely, one of them on a NW sidewall facet and the other one on a NW top facet. Similarly to the low-temperature PL data shown in Fig.~\ref{fig:LTPL}, we observe the near-band edge emission and the green band luminescence in all cases. The transitions are, however, shifted towards longer wavelengths due to the temperature-induced reduction in band-gap. As in the case of the PL, the integrated intensity of the NW CL spectra is higher than that of the layer. Interestingly, when comparing the two NW CL spectra, we find a remarkable difference. As can be observed, the intensity ratio between the near-band-edge emission and the defect-related green luminescence is much higher for the spectrum acquired at the NW top facet. 

\begin{figure*}
\hspace{1in}\includegraphics*[width=0.85\textwidth]{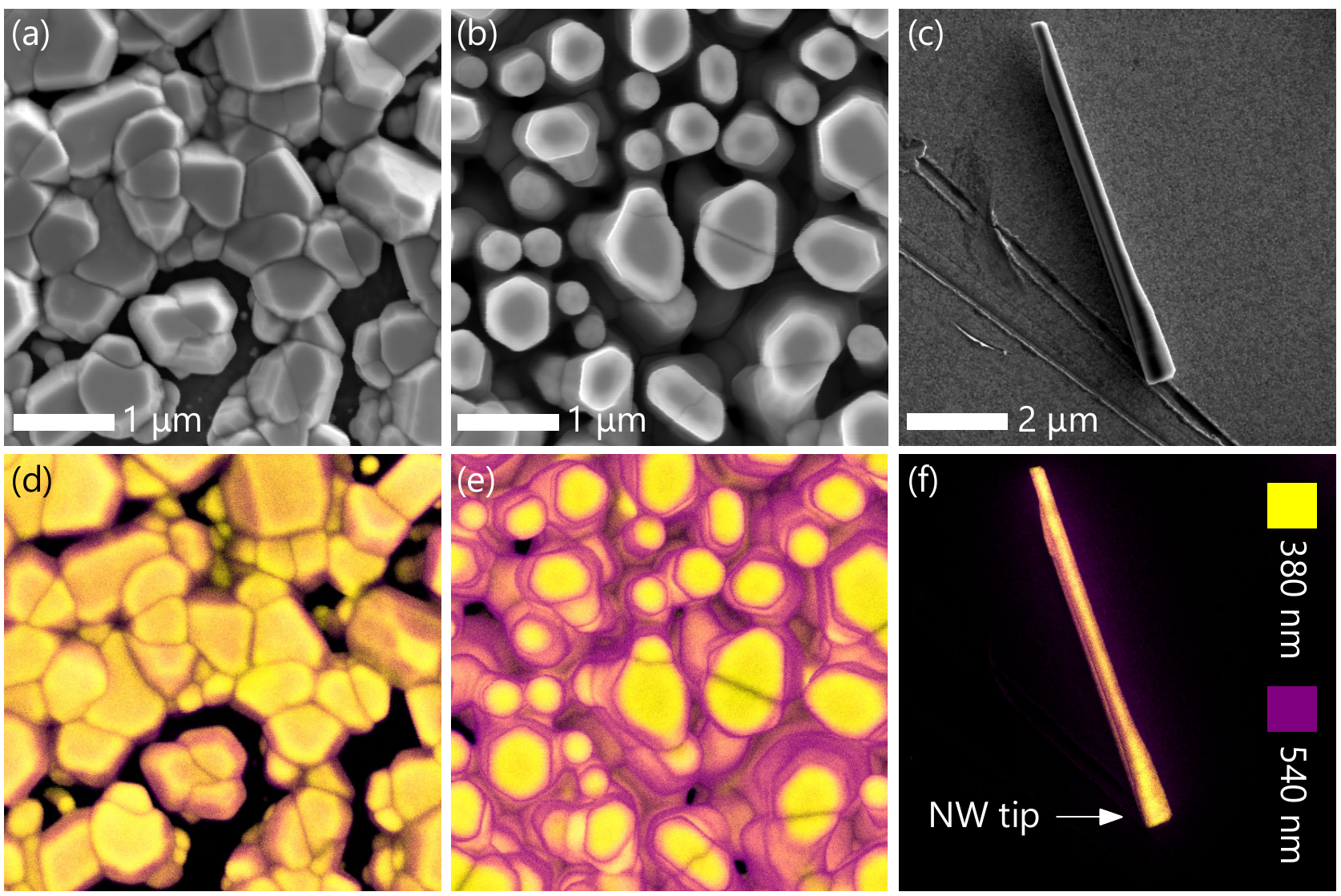}
\caption{\label{fig:CL_maps}(Colour online) Top panels: plan-view scanning electron micrographs of (a) the pseudo-compact layer, (b) the NW ensemble, and (c) a single NW mechanically transferred onto a Si wafer. Bottom panels: corresponding superimposed false-colour CL intensity maps at 380 (yellow) and 540~nm (purple), as indicated in (f).}
\end{figure*}

To gain further insights into the correlation between the morphology and luminescence properties of the sample on a microscopic scale, it is simultaneously imaged by SEM and CL at room-temperature. Figure~\ref{fig:CL_maps} shows plan-view scanning electron micrographs together with their corresponding false-colour CL intensity maps recorded at emission wavelengths of 380 and 540~nm superimposed on each other for the pseudo-compact layer, the NW ensemble and a representative single NW mechanically transferred onto a Si wafer. In the case of the pseudo-compact layer, the near-band-edge emission mainly arises from the top surface of the three-dimensional islands and  it is particularly quenched at coalescence joints. In contrast, the green luminescence is observed everywhere. This emission is clearly enhanced at coalescence joints as well as at the non-polar and semi-polar facets formed at the edges of the three dimensional islands. The luminescence distribution of the NW ensemble is slightly different from that of the layer. In analogy to the former case, the near-band-edge emission comes from the top facets of the NWs and quenches at coalescence joints. However, in contrast to the layer, the green luminescence predominantly arises from the sidewall facets of the NWs. Importantly, the green emission is not always present at the coalescence boundaries. This result, together with the lack of near-band-edge luminescence at these locations, evidences the formation of a higher concentration of non-radiative recombination centres at the joints between adjacent NWs. The analysis of the single NW dispersed on the Si wafer indicates that, even though the near-band-edge emission is stronger close to the NW tip, it is generated in the core of the NW along its entire length. On the contrary, the green luminescence primarily arises from the volume close to the NW sidewall facets, as also observed in Refs.~\cite{Nobis2004,Djurisic2006,Foley2008,Xue2009,Fabbri2015,Zhu2015,Ruane2016,Yao2016} for ZnO NWs grown by different techniques. 

\begin{figure*}
\includegraphics*[width=\textwidth]{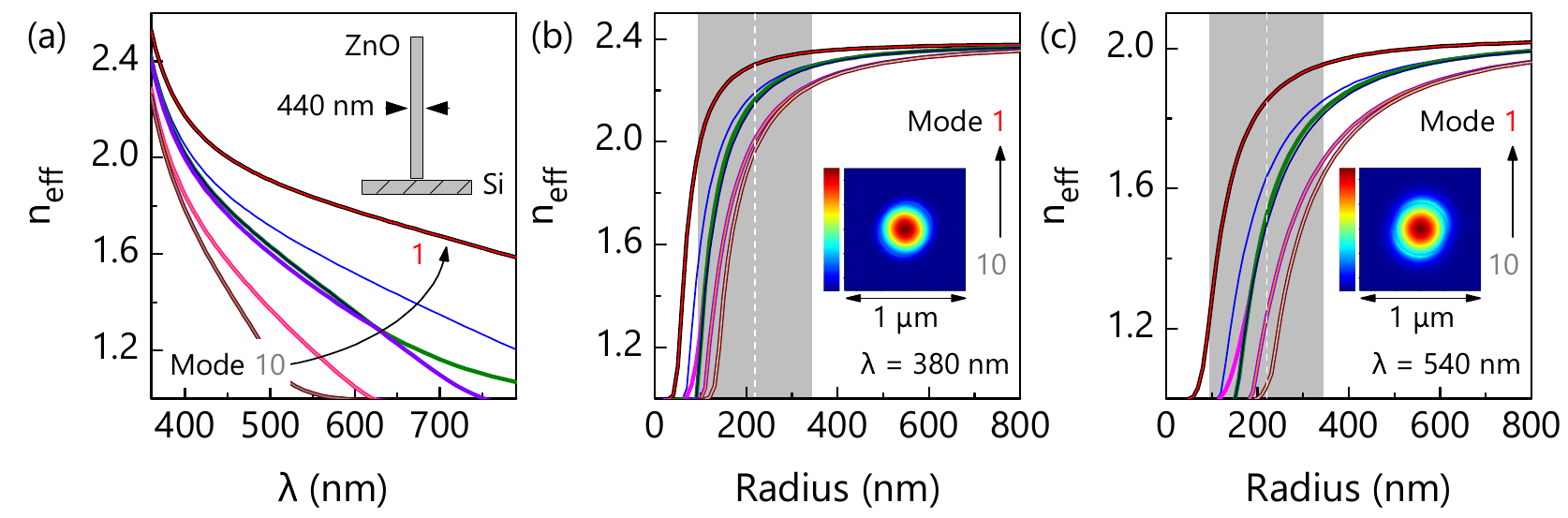}
\caption{\label{fig:FEEM}(Colour online) (a) Effective refractive index for the first ten guided modes as a function of the light wavelength for a given NW radius of 220~nm, i.\,e., the mean NW radius derived from the shifted gamma function shown in Fig.~\ref{circularity}(c). The inset illustrates the simulated structure, a single upright NW surrounded by air on a Si wafer. (b) and (c) Effective refractive index for the first ten guided modes as a function of the NW radius for (b) $\lambda=380$~nm and (c) $\lambda=540$~nm. The dashed vertical lines indicate the mean NW radius (220~nm) and the grey shadowed areas the width of the NW radius distribution. The insets in (b) and (c) show the cross-sectional intensity map of the electric field for the fundamental guided mode of a NW with a radius of 220~nm at 380 and 540~nm, respectively. The intensity is represented in arbitrary units using the linear colour code shown by the scale bars next to the plots.}
\end{figure*}

The different spatial distributions of the near-band-edge emission and the defect-related green luminescence may be caused by either a comparatively high concentration of point defects at non-polar and semi-polar facets \cite{Nobis2004,Djurisic2006,Foley2008,Xue2009,Rodnyi2011,Fabbri2015,Zhu2015,Ruane2016,Yao2016} or waveguiding effects. Due to the high refractive index of ZnO of about 2 between 380 and 700~nm, ZnO NWs could support waveguide modes, depending on both the NW radius \textit{R} and the wavelength $\lambda$ of the light \cite{Voss2007,Brenny2016,Hauswald2017,Verardo2018}. The different luminescence distributions at 380 and 540~nm could thus be explained in terms of waveguiding effects, provided that there are waveguide modes for the former wavelength, but not for the latter. To investigate the existence of waveguide modes in our ZnO NWs, we perform two dimensional simulations using the software Lumerical FEEM, a finite element Maxwell equation solver based on the eigenmode method. For the simulations, we take into account the wavelength dependence of the ZnO refractive index reported in Ref.~\cite{Oezguer2006}. The inset of Fig.~\ref{fig:FEEM}(a) illustrates the simulated structure, a single, upright, cylindrical ZnO NW surrounded by air on a Si wafer. Figure~\ref{fig:FEEM}(a) shows the wavelength dependence of the effective refractive index \cite{Selvaraja2018} for the first ten modes as a function of the wavelength for the mean NW radius derived from Fig.~\ref{histograms}(c), i.\,e., 220~nm. Although the effective refractive index for the different modes rapidly decreases with increasing values of $\lambda$ for the two wavelengths of interest, 380 and 540~nm, all ten modes are guided since $n_{eff}>1$. In fact, for the fundamental mode at these two wavelengths, the electric field distribution along the NW cross section is rather similar, as shown in the insets of Figs.~\ref{fig:FEEM}(b) and \ref{fig:FEEM}(c). Consequently, there are no significant differences at 380 and 540~nm, at least for those NWs with a radius close to 220~nm. To examine the impact of the NW radius on the guided modes, we compute the variation of the effective refraction index with \textit{R} for $\lambda=380$~nm and $\lambda=540$~nm. Figures~\ref{fig:FEEM}(b) and \ref{fig:FEEM}(c) present the respective results for the first ten modes. In both figures, the grey shadowed area indicates the width of the above determined NW radius distribution. As expected, regardless of the wavelength, a NW with a larger radius exhibits a higher number of guided modes, with their effective refractive index increasing monotonically with increasing values of \textit{R}. For $\lambda=380$~nm, the thinnest NWs ($R=95$~nm) support four guided modes, and those with $R>125$~nm the first ten modes. For $\lambda=540$~nm, only the first two modes are guided for $R=95$~nm, and the NW radius must be larger than 210~nm to support the first ten modes. Hence, if a significant amount of radiation couples into high-order modes, we should observe variations in the luminescence distribution depending on the NW radius. The inspection of Fig.~\ref{fig:CL_maps}(e) demonstrates that it is not the case, the luminescence distribution is identical for thin and thick NWs. Consequently, for both wavelengths, most of the vertically guided radiation is primarily coupled into low-order modes. The differences observed in the luminescence distribution at 380 and 540~nm can thus only be explained if the green-emitting radiative defects were not homogeneously distributed, but located at or close to the NW sidewalls, as previously reported \cite{Nobis2004,Djurisic2006,Foley2008,Xue2009,Fabbri2015,Zhu2015,Ruane2016,Yao2016}. In such a situation, the green radiation would barely couple into the guided modes, instead leaking out through the lateral surfaces in all directions, as seen in Figs.~\ref{fig:CL_maps}(e) and \ref{fig:CL_maps}(f). Based on these arguments, we conclude that the NWs exhibit a core-shell structure, where the shell is characterized by a high concentration of green-emitting radiative point defects with respect to the core. According to this interpretation, the more homogeneous green-luminescence distribution observed in the pseudo-compact layer as compared to the NW ensemble [see Figs.~\ref{fig:CL_maps}(d) and \ref{fig:CL_maps}(e)] suggests that the incorporation of green-emitting radiative point defects is either enhanced during radial growth or favoured under the specific growth conditions promoting the formation of ZnO layers in CVT. 

\section{Discussion}

The examination of the morphological properties of the NW ensemble revealed a non-negligible degree of coalescence that can be quantitatively assessed by analysing the cross-sectional shapes of the NWs. Due to the mutual misorientation of coalesced NWs both out of plane and in plane, a large density of structural defects are expected to form at the coalescence joints, in particular dislocated tilt and twist boundaries. The analysis of the ensemble by CL shows that coalesced joints act as sources of non-radiative recombination. Therefore, to maximize the potential of self-assembled ZnO NWs grown on Si by CVT for optoelectronic applications, the origin of coalescence must be identified and eliminated. 

In the present case, there are three potential sources of coalescence: (i) NWs could coalesce due to their out-of-plane misorientation. According to the analysis of the sample by XRD, the NWs are randomly tilted with a distribution characterized by a width of about $6^{\circ}$. Considering that the mean separation between NWs is approximately 250~nm, as follows from the number density and radius distribution of the NW ensemble, two adjacent NWs with a tilt of $6^{\circ}$ will merge together upon reaching a length of $\approx1.2$~$\mu$m. Hence, in the present case, where the NWs are several micrometers long, NW tilt clearly is a non-negligible source of coalescence. (ii) NWs could also merge together due to their radial growth, a phenomenon favoured by their apparent random nucleation on Si. (iii) Last but not least, coalescence could also be caused by NW bundling \cite{Han2006,Liu2008,Zhu2015,Kaganer_NL_2016}, a process promoted by the reduction of surface energy at the expense of the elastic energy of bending \cite{Kaganer_NL_2016}. In other words, NWs bend and merge together because, above a certain critical length, the reduction in the total surface area associated to the formation of coalesced aggregates makes the process energetically favourable. This source of coalescence, presumably triggered by electrostatic forces \cite{Liu2008,Zhu2015}, has been previously observed in ZnO as well as in other material systems \cite{Han2006,Liu2008,Khorasaninejad2012,Chen2013,Sun2014,Carapezzi2014,Zhu2015,Kaganer_NL_2016} and may be related to the apparent preferential formation of coalescence boundaries parallel to \textit{A}- and \textit{M}-planes, a rather unexpected result taking into account the random in-plane orientation distribution of the NWs. Regardless of the dominant source of coalescence, the formation of NW aggregates will always be minimized by using growth conditions and/or nucleation layers resulting in lower NW number densities. In the particular case of the coalescence induced by tilt, considering that ZnO NWs tend to elongate perpendicularly to the substrate surface, coalescence events can be further reduced by decreasing, as much as possible, the roughness of the surface on which the NWs nucleate. Finally, besides reducing the NW density and tilt, since for processes (i) and (iii) coalescence sets in only upon exceeding certain critical lengths, these two sources of coalescence can be totally suppressed by limiting the NW length. 

\section{Conclusions}
Self-assembled ZnO NWs grown by chemical vapour transport on Si(001) using a Zn layer to enhance their nucleation crystallize in the wurtzite phase and elongate along the \textit{c}-axis, as typically observed for this type of nanostructures. For the substrate preparation and growth conditions employed in this study, the NWs exhibit a tilt distribution with a width of approximately $6^{\circ}$ and are randomly oriented in plane. The luminescence of the NWs is characterized by two bands, the near-band-edge luminescence, attributed to the presence of various types of donors, and the well-known defect-related green luminescence of ZnO, associated to Cu impurities and native point defects. Our study indicates that, during growth, the point defects responsible for the green luminescence segregate toward non-polar and semi-polar facets. Hence, the green luminescence of our ZnO NWs primarily arises from their sidewall facets, and could thus be eventually mitigated by appropriate surface treatments. Importantly, the formation of coalesced NW aggregates during growth does not only distort the shape and dimensions of ZnO NWs, but also severely damages their luminescence efficiency due to the creation of non-radiative recombination centres at coalescence joints. Therefore, the coalescence degree is an important parameter that can be used as a figure of merit to assess the quality of ZnO NW ensembles. We demonstrate that, as in the case of GaN, the coalescence degree can be unambiguously quantified mathematically by analysing either the circularity or the area-perimeter plot of the cross-sectional shape of the NWs. In the former case, our analysis evidences that, for ensembles of ZnO NWs randomly oriented in plane, a suitable threshold value to distinguish between single NWs and coalesced aggregates is 0.65. The possibility of a quantitative assessment of the coalescence degree of ZnO NW ensembles paves the way for future studies devoted to systematically correlate this parameter with either the growth conditions or different physical properties, such as the luminescence efficiency and the strain state of ZnO NWs.

\ack 

We are indebted to Eduardo Ruíz for his technical support during the preparation of the sample and grateful to Oliver Brandt for  discussions on the physical origin of the inhomogeneous luminescence distribution in ZnO NWs. S. Fernández-Garrido and A. Redondo-Cubero gratefully acknowledge the final support received through the Spanish program Ramón y Cajal (co-financed by the European Social Fund) under grants RYC-2016-19509 and RYC-2015-18047, respectively, from the former Ministerio de Ciencia, Innovación y Universidades. We also acknowledge the funding provided by the projects CTQ-2017-84309-C2-2-R from MINECO and P2018/NMT4349 from Comunidad Autónoma de Madrid.


\bibliographystyle{iopart-num}

\bibliography{bibliography}

\providecommand{\newblock}{}
\begin{thebibliography}{10}
\expandafter\ifx\csname url\endcsname\relax
  \def\url#1{{\tt #1}}\fi
\expandafter\ifx\csname urlprefix\endcsname\relax\def\urlprefix{URL }\fi
\providecommand{\eprint}[2][]{\url{#2}}

\bibitem{Patolsky2005}
Patolsky F and Lieber C~M 2005 {\em Mater. Today\/} {\bf 8} 20

\bibitem{Li2006}
Li Y, Qian F, Xiang J and Lieber C~M 2006 {\em Mater. Today\/} {\bf 9} 18

\bibitem{Yan2009}
Yan R, Gargas D and Yang P 2009 {\em Nat. Photonics\/} {\bf 3} 569

\bibitem{Yi2005}
Yi G~C, Wang C and Park W~I 2005 {\em Semicond. Sci. Technol.\/} {\bf 20}

\bibitem{Schmidt-mende_2007}
Schmidt-Mende L and Macmanus-Driscoll J~L 2007 {\em Mater. Today\/} {\bf 10} 40

\bibitem{Xu2011}
Xu S and Wang Z~L 2011 {\em Nano Res.\/} {\bf 4} 1013

\bibitem{Consonni_PSSRL_2013}
Consonni V 2013 {\em Phys. Status Solidi RRL\/} {\bf 7} 699

\bibitem{Calabrese_APL_2016}
Calabrese G, Corfdir P, Gao G, Pf{\"{u}}ller C, Trampert A, Brandt O, Geelhaar
  L and Fern{\'{a}}ndez-Garrido S 2016 {\em Appl. Phys. Lett.\/} {\bf 108}
  202101

\bibitem{LePivert2019}
{Le Pivert} M, Poupart R, Capochichi-Gnambodoe M, Martin N and Leprince-Wang Y
  2019 {\em Microsyst Nanoeng\/} {\bf 5} 57

\bibitem{Campos2020}
Campos A~C, Paes S~C, Correa B~S, Cabrera-Pasca G~A, Costa M~S, Costa C~S,
  Otubo L and Carbonari A~W 2020 {\em ACS Appl. Nano Mater.\/} {\bf 3} 175

\bibitem{Wagner_apl_1964}
Wagner R~S and Ellis W~C 1964 {\em Appl. Phys. Lett.\/} {\bf 4} 89

\bibitem{Comini2016}
Comini E 2016 {\em Mater. Today\/} {\bf 19} 559

\bibitem{Shanmugam2017}
Shanmugam N~R, Muthukumar S and Prasad S 2017 {\em Future Science OA\/} {\bf 3}
  FSO196

\bibitem{Djurisic2006}
Djuri{\v{s}}i{\'{c}} A~B and Leung Y~H 2006 {\em Small\/} {\bf 2} 944

\bibitem{Reshchikov2009}
Reshchikov M~A, Behrends A, Bakin A and Waag A 2009 {\em J. Vac. Sci. Technol.
  B\/} {\bf 27} 1688

\bibitem{Zettler_CGD_2015}
Zettler J~K, Hauswald C, Corfdir P, Musolino M, Geelhaar L, Riechert H, Brandt
  O and Fern{\'{a}}ndez-Garrido S 2015 {\em Cryst. Growth Des.\/} {\bf 15} 4104

\bibitem{Consonni_APL_2011_2}
Consonni V, Knelangen M, Trampert A, Geelhaar L and Riechert H 2011 {\em Appl.
  Phys. Lett.\/} {\bf 98} 071913

\bibitem{Jenichen_NT_2011}
Jenichen B, Brandt O, Pf{\"{u}}ller C, Dogan P, Knelangen M and Trampert A 2011
  {\em Nanotechnology\/} {\bf 22} 295714

\bibitem{Fan2014}
Fan S, Zhao S, Liu X and Mi Z 2014 {\em J. Vac. Sci. Technol. B\/} {\bf 32}
  02C114

\bibitem{Brandt_CGD_2014}
Brandt O, Fern{\'{a}}ndez-Garrido S, Zettler J~K, Luna E, Jahn U, Ch{\`{e}}ze C
  and Kaganer V~M 2014 {\em Cryst. Growth Des.\/} {\bf 14} 2246

\bibitem{Fernandez-Garrido_NT_2014}
Fern\'{a}ndez-Garrido S, Kaganer V~M, Hauswald C, Jenichen B, Ramsteiner M,
  Consonni V, Geelhaar L and Brandt O 2014 {\em Nanotechnology\/} {\bf 25}
  455702

\bibitem{Kaganer_NL_2016}
Kaganer V~M, Fern\'andez-Garrido S, Dogan P, Sabelfeld K~K and Brandt O 2016
  {\em Nano Lett.\/} {\bf 16} 3717

\bibitem{Auzelle2016}
Auzelle T, Biquard X, Bellet-Amalric E, Fang Z, Roussel H, Cros A and Daudin B
  2016 {\em J. Appl. Phys.\/} {\bf 120} 225701

\bibitem{Treeck2018}
van Treeck D, Calabrese G, Goertz J~J~W, Kaganer V~M, Brandt O,
  Fern{\'{a}}ndez-Garrido S and Geelhaar L 2018 {\em Nano Res.\/} {\bf 11} 565

\bibitem{Consonni2009}
Consonni V, Knelangen M, Jahn U, Trampert A, Geelhaar L and Riechert H 2009
  {\em Appl. Phys. Lett.\/} {\bf 95} 241910

\bibitem{Grossklaus_JCG_2013}
Grossklaus K~A, Banerjee A, Jahangir S, Bhattacharya P and Millunchick J~M 2013
  {\em J. Cryst. Growth\/} {\bf 371} 142

\bibitem{Jeong2005}
Jeong J~S, Lee J~Y, Cho J~H, Suh H~J and Lee C~J 2005 {\em Chem. Mater.\/} {\bf
  17} 2752

\bibitem{Han2006}
Han X, Wang G, Zhou L and Hou J~G 2006 {\em Chem. Commun.\/}  212

\bibitem{Liu2008}
Liu J, Lee S, Lee K, Ahn Y~H, Park J~Y and Koh K~H 2008 {\em Nanotechnology\/}
  {\bf 19} 185607

\bibitem{Kumar2013}
Kumar E~S, Anderson I~P, Deng Z, Mohammadbeigi F, Wintschel T, Huang D and
  Watkins S~P 2013 {\em Semicond. Sci. Technol.\/} {\bf 28} 045014

\bibitem{Zhu2015}
Zhu L, Phillips M~R and Ton-That C 2015 {\em Cryst. Eng. Comm.\/} {\bf 17} 4987

\bibitem{Fu1998}
Fu Z, Lin B, Liao G and Wu Z 1998 {\em J. Cryst. Growth\/} {\bf 193} 316

\bibitem{GarciaNunez2014}
{Garc{\'{i}}a N{\'{u}}{\~{n}}ez} C, Pau J, Ru{\'{i}}z E, {Garc{\'{i}}a
  Mar{\'{i}}n} A, Garc{\'{i}}a B, Piqueras J, Shen G, Wilbert D, Kim S and Kung
  P 2014 {\em Thin Solid Films\/} {\bf 555} 42

\bibitem{Park2008}
Park W~I 2008 {\em Met. Mater. Int.\/} {\bf 14} 659

\bibitem{Abramoff_Biophotonics_2004}
{Abramoff, MD; Magalh\~{a}es, Paulo J; Ram} S~J 2004 {\em Biophotonics Int.\/}
  {\bf 11} 36

\bibitem{Meyer2004}
Meyer B~K, Alves H, Hofmann D~M, Kriegseis W, Forster D, Bertram F, Christen J,
  Hoffmann A, Stra{\ss}burg M, Dworzak M, Haboeck U and Rodina A~V 2004 {\em
  Phys. Status Solidi B\/} {\bf 241} 231

\bibitem{Oezguer2006}
\"{O}zg\"{u}r U and Morko\c{c} H 2006 Chapter 5 {\em Zinc Oxide Bulk, Thin
  Films and Nanostructures\/} ed Jagadish C and Pearton S (Oxford: Elsevier
  Science Ltd) pp 175 -- 239

\bibitem{ZYao2015}
Yao Z, Gu S, Tang K, Ye J, Zhang Y, Zhu S and Zheng Y 2015 {\em J. Lumin.\/}
  {\bf 161} 293

\bibitem{Dingle1969}
Dingle R 1969 {\em Phys. Rev. Lett.\/} {\bf 23} 579

\bibitem{Reshchikov2007}
Reshchikov M, Morko{\c{c}} H, Nemeth B, Nause J, Xie J, Hertog B and Osinsky A
  2007 {\em Physica B\/} {\bf 401-402} 358

\bibitem{Katiyar2013}
Katiyar R~S and Samanta K 2013 Handbook of zinc oxide and related materials -
  volume one {\em Structural and optical properties of
  Zn$_\mathrm{1-x}$Cu$_\mathrm{x}$O thin films\/} ed Feng Z~C (CRC Press -
  Taylor and Francis Group) pp 351 -- 372

\bibitem{Park2010}
Park H, Byeon K~J, Yang K~Y, Cho J~Y and Lee H 2010 {\em Nanotechnology\/} {\bf
  21} 355304

\bibitem{Djavid2016}
Djavid M and Mi Z 2016 {\em Appl. Phys. Lett.\/} {\bf 108} 051102

\bibitem{Hauswald2017}
Hauswald C, Giuntoni I, Flissikowski T, Gotschke T, Calarco R, Grahn H~T,
  Geelhaar L and Brandt O 2017 {\em ACS Photonics\/} {\bf 4} 52

\bibitem{Lin2018}
Lin R, Galan S~V, Sun H, Hu Y, Alias M~S, Janjua B, Ng T~K, Ooi B~S and Li X
  2018 {\em Photonics Res.\/} {\bf 6} 457

\bibitem{Nobis2004}
Nobis T, Kaidashev E~M, Rahm A, Lorenz M, Lenzner J and Grundmann M 2004 {\em
  Nano Lett.\/} {\bf 4} 797

\bibitem{Foley2008}
Foley M, Ton-That C and Phillips M~R 2008 {\em Appl. Phys. Lett.\/} {\bf 93}
  243104

\bibitem{Xue2009}
Xue H, Pan N, Zeng R, Li M, Sun X, Ding Z, Wang X and Hou J~G 2009 {\em J.
  Phys. Chem. C\/} {\bf 113} 12715

\bibitem{Fabbri2015}
Fabbri F, Villani M, Catellani A, Calzolari A, Cicero G, Calestani D, Calestani
  G, Zappettini A, Dierre B, Sekiguchi T and Salviati G 2015 {\em Sci. Rep.\/}
  {\bf 4} 5158

\bibitem{Ruane2016}
Ruane W~T, Johansen K~M, Leedy K~D, Look D~C, von Wenckstern H, Grundmann M,
  Farlow G~C and Brillson L~J 2016 {\em Nanoscale\/} {\bf 8} 7631

\bibitem{Yao2016}
Yao Z, Tang K, Xu Z, Ye J, Zhu S and Gu S 2016 {\em Nanoscale Res. Lett.\/}
  {\bf 11} 511

\bibitem{Rodnyi2011}
Rodnyi P~A and Khodyuk I~V 2011 {\em Optics and Spectroscopy\/} {\bf 111} 776

\bibitem{Voss2007}
Voss T, Svacha G~T, Mazur E, M{\"{u}}ller S, Ronning C, Konjhodzic D and Marlow
  F 2007 {\em Nano Lett.\/} {\bf 7} 3675

\bibitem{Brenny2016}
Brenny B~J~M, Abujetas D~R, van Dam D, S{\'{a}}nchez-Gil J~A, Rivas J~G and
  Polman A 2016 {\em ACS Photonics\/} {\bf 3} 677

\bibitem{Verardo2018}
Verardo D, Lindberg F~W, Anttu N, Niman C~S, Lard M, Dabkowska A~P, Nylander T,
  M{\aa}nsson A, Prinz C~N and Linke H 2018 {\em Nano Lett.\/} {\bf 18} 4796

\bibitem{Selvaraja2018}
Selvaraja S~K and Sethi P 2018 {Review on Optical Waveguides} {\em Emerging
  Waveguide Technology\/} (InTech) pp 95--129

\bibitem{Khorasaninejad2012}
Khorasaninejad M, Abedzadeh N, {Singh Jawanda} A, O N, Anantram M~P and {Singh
  Saini} S 2012 {\em J. Appl. Phys.\/} {\bf 111} 044328

\bibitem{Chen2013}
Chen B, Gao Q, Chang L, Wang Y, Chen Z, Liao X, Tan H~H, Zou J, Ringer S~P and
  Jagadish C 2013 {\em Acta Mater.\/} {\bf 61} 7166

\bibitem{Sun2014}
Sun Z, Wang D and Xiang J 2014 {\em ACS Nano\/} {\bf 8} 11261

\bibitem{Carapezzi2014}
Carapezzi S, Priante G, Grillo V, Mont{\`{e}}s L, Rubini S and Cavallini A 2014
  {\em ACS Nano\/} {\bf 8} 8932

\end{thebibliography}

\end{document}